\documentclass[letter,english,superscriptaddress,twocolumn,prb,noshowkeyws,showpacs]{revtex4-1}

\usepackage[T1]{fontenc}
\usepackage[latin9]{inputenc}
\usepackage{amsmath}
\usepackage{graphicx}
\usepackage{amssymb}
\usepackage{esint}
\usepackage{setspace}
\usepackage{babel}
\usepackage[normalem]{ulem}
\usepackage{amsmath,amstext,amsbsy,amsopn,amsthm,amscd,amsfonts,amssymb}
\usepackage{bm}
\usepackage{epsfig}
\usepackage{graphpap}
\usepackage{mathrsfs}
\usepackage{setspace}
\usepackage{subfigure}

\theoremstyle{definition}

\theoremstyle{remark}

\newcommand{\REMOVE}[1]{}

\newcommand{\begeq}{\begin{equation}}
\newcommand{\eneq}{\end{equation}}
\newcommand{\beq}{\begin{equation}}
\newcommand{\eeq}{\end{equation}}
\newcommand{\beqa}{\begin{eqnarray}}
\newcommand{\eeqa}{\end{eqnarray}}

\newcommand{\eps}{\epsilon}
\newcommand{\um}{{\mu{m}}}
\newcommand{\fig}[1]{Fig. \ref{#1}}
\newcommand{\BSCCO}{Bi$_2$Sr$_2$CaCu$_2$O$_{8+\delta}$~}
\newcommand{\tild}{\tilde{d}}

\begin{document}

\title{Suppression of geometrical barrier in \BSCCO crystals by Josephson vortex stacks}

\author{Y. Segev}
\email{yehonathan.segev@gmail.com}
\homepage{http://www.weizmann.ac.il/condmat/superc/}
\affiliation{Department of Condensed Matter Physics, Weizmann Institute of Science, Rehovot 76100, Israel}
\author{I. Gutman}
\affiliation{Department of Condensed Matter Physics, Weizmann Institute of Science, Rehovot 76100, Israel}

\author{S. Goldberg}
\affiliation{Physics Department, Duke University, Durham, NC 27708,
USA}

\author{Y. Myasoedov}
\affiliation{Department of Condensed Matter Physics, Weizmann Institute of Science, Rehovot 76100, Israel}

\author{E. Zeldov}
\affiliation{Department of Condensed Matter Physics, Weizmann Institute of Science, Rehovot 76100, Israel}

\author{E. H. Brandt}
\affiliation{Max-Planck-Institut f\"{u}r Metallforschung, D-70506 Stuttgart, Germany}

\author{G. P. Mikitik}
\affiliation{B. Verkin Institute for Low Temperature Physics \& Engineering, National Ukrainian Academy of Sciences, Kharkov 61103, Ukraine} %
\affiliation{Max-Planck-Institut f\"{u}r Metallforschung, D-70506 Stuttgart, Germany}

\author{T. Katagiri}
\affiliation{Materials and Structures Laboratory, Tokyo Institute of Technology, Kanagawa 226-8503,
Japan}

\author{T. Sasagawa}
\affiliation{Materials and Structures Laboratory, Tokyo Institute of Technology, Kanagawa 226-8503,
Japan}

\begin{abstract}
Differential  magneto-optics are used to study the effect
of dc in-plane magnetic field on hysteretic behavior due to
geometrical barriers in \BSCCO crystals. In the absence of
an in-plane field a vortex dome is visualized in the sample
center surrounded by barrier-dominated flux-free regions.
With an in-plane field, stacks of Josephson vortices form
vortex chains which are surprisingly found to protrude out
of the dome and into {the} vortex-free regions. The chains
are imaged to extend up to the sample edges, thus providing
easy channels for vortex entry and for drain of the dome
through geometrical barrier, suppressing the magnetic
hysteresis. Reduction of the vortex energy due to crossing
with Josephson vortices is evaluated to be about two orders
of magnitude too small to account for the formation of the
protruding chains. We present a model and numerical
calculations that qualitatively describe the observed
phenomena by taking into account the demagnetization
effects in which flux expulsion from the pristine regions
results in vortex focusing and in the chain protrusion.
Comparative measurements on a sample with narrow etched
grooves provide further support to the proposed model.
\end{abstract}
\maketitle

\section{introduction}

One of the most prominent characteristics of type II
superconductors is the large magnetic hysteresis that
arises from the existence of potential barriers that impede
vortex motion.\cite{Blatter-1994} There are several sources
of potential barriers: pinning due to quenched material
disorder in the bulk of the sample, surface pinning
produced by the roughness of the sample
surface,\cite{Pautrat-2004} Bean-Livingston surface barrier
arising from the force between a vortex parallel to a
surface and its image,\cite{Bean-1964} and geometrical
barrier (GB) that arises from the force due to Meissner
currents in thin slab-shaped superconducting samples in
perpendicular
field.\cite{Zeldov-1994,Indenbom-1994,{Benkraouda-1996},Brandt-1999,Brandt-1999a,Schuster-1994,Clem-2008}

Strong vortex pinning and the associated magnetic
hysteresis are vital elements in superconductor
applications that are based on high critical currents. The
magnetic hysteresis, however, masks the equilibrium
magnetization properties, the study of which is essential
for comprehension of the thermodynamic structure of the
vortex matter. In order to overcome this limitation, vortex
shaking has been employed in recent years, which was shown
to suppress the magnetic hysteresis very effectively at not
too low temperatures.\cite{Willemin-1998, Avraham-2001,
Beidenkopf-2005} In this method an ac in-plane field
$H_{x}$ is applied in addition to the main dc $H_z$ which
causes local shaking and equilibration of the vortices. dc
in-plane fields were also shown to reduce the magnetic
hysteresis.\cite{Tamegai-2004a,Bending-2004,Kasahara-2005,Konczykowski-2006,Gutman-2009}

Vortex shaking was found to suppress the hysteresis that
arises either from bulk pinning that is usually the
dominant source of hysteresis at lower temperatures,
surface barriers that often dominate vortex dynamics at
elevated temperatures and
fields,\cite{Majer-1995,VanderBeek-1996,Fuchs-1998,
Fuchs-1998b, Beidenkopf-2009, Bohmer-2010} and GB's that
commonly govern the hysteresis at high temperatures and low
fields.\cite{Zeldov-1994,Indenbom-1994,Majer-1995,Morozov-1997,Bohmer-2010}
Detailed theoretical studies explained the suppression of
bulk pinning hysteresis by vortex shaking in terms of the
Bean model in a turning magnetic
field.\cite{Brandt-2002,Mikitik-2003,Mikitik-2004} This
description may similarly apply to the case of surface
pinning. Also the suppression of the Bean-Livingston
surface barrier hysteresis could possibly be understood in
terms of easier vortex loop nucleation at the sample
corners for tilted vortices.\cite{Burlachkov-1994} In the
above mechanisms the potential barriers are microscopic and
hence local ac displacement or tilting of a vortex segment
in the presence of an average dc driving force can enhance
vortex activation and relaxation. In contrast, the GB in a
platelet sample gives rise to a vortex dome in the center
of the sample surrounded by a vortex-free
region.\cite{Zeldov-1994,Indenbom-1994,Benkraouda-1996,Brandt-1999,Brandt-1999a}
This vortex-free region constitutes a large barrier that
has a macroscopic energy scale of the order of the vortex
line tension times the sample thickness and a macroscopic
length scale of the order of sample's width. As a result,
this barrier cannot be overcome by any local vortex
perturbation or tilting. Hence the experimental observation
that the vortex shaking by the  in-plane field suppresses
the hysteresis due to GB was highly surprising and has
remained unresolved so far.

In this paper we describe a possible resolution of this
question in the case of layered superconductors by
investigating the GB behavior in the presence of a dc
in-plane field. In highly anisotropic layered materials
like \BSCCO \; (BSCCO) the in-plane field $H_{x}$ results
in {the} formation of stacks of Josephson vortices
{(JV's)}.\cite{Clem-1991,Bulaevskii-1992,Koshelev-1999} The
addition of $H_z$ causes the formation of crossing lattices
in which stacks of pancake vortices (PV's) form chains
along the Josephson
vortices.\cite{Koshelev-1999,Grigorenko-2001,Bending-2005,Tokunaga-2003,Bolle-1991,Vlasko-Vlasov-2002,Tamegai-2004}
Surprisingly, we find that these chains exist also outside
the vortex dome, protruding like vortex whiskers into
regions that are vortex free in the absence of $H_{x}$. As
a result, easy channels are apparently formed that allow
vortex penetration through the macroscopic GB suppressing
the hysteresis. This observation appears to be inconsistent
with theoretical expectations since the reduction in the
energy of the pancake stack by the Josephson vortices
should be much smaller than the height of the geometrical
barrier.\cite{Zeldov-1994,Koshelev-1999} Therefore, the
regions outside the vortex dome should have remained vortex
free even in the presence of $H_{x}$ and no significant
suppression of the barrier should occur. We propose an
explanation to this phenomenon in terms of interaction
between JV's and PV's that is enhanced by the
demagnetization effects in platelet geometry. We
demonstrate this scenario by studying vortex behavior in a
sample with a shallow and narrow groove etched into its
surface using a focused ion beam (FIB).


\section {geometrical barrier}
\label{background} The GB is a macroscopic potential barrier that
vortices have to overcome in order to enter or leave the
sample and is the main source of magnetic hysteresis in
thin slab-shaped samples with weak pinning. The barrier is
formed by the interplay between the vortex line tension and
the Lorentz force that is induced by the circulating
Meissner
currents.\cite{Zeldov-1994,Indenbom-1994,Benkraouda-1996,Brandt-1999,Brandt-1999a,Schuster-1994,Clem-2008}
In a sample with elliptical cross section, for example, the
energy of a test vortex has two contributions: the positive
vortex line energy increases gradually from zero at the
edge of the sample to a value of $\eps_0 d$ in the center
($\eps_0$ is the vortex line energy per unit length and $d$
is the sample thickness). The Meissner currents that flow
over the entire surface exert inward forces on the vortex
that gradually lower its energy as it moves toward the
center. In a sample with elliptical cross section the two
contributions cancel out rendering a constant vortex energy
and a corresponding position independent and reversible
induction $B(x,H)$. In a platelet sample of rectangular
cross section, in contrast, by cutting through the sharp
rims, the vortex attains its full line energy $\eps_0 d$
within a distance of the order of $d/2$ from the
edge,\cite{Zeldov-1994} while the Meissner currents remain
distributed as in elliptical sample.\cite{Zeldov-1994b} As
a result, a geometry related barrier of height of the order
of $\eps_0 d$ is formed that extends over a width of the
order of the half width of the sample,
$w$.\cite{Zeldov-1994,Indenbom-1994,Benkraouda-1996,Brandt-1999,Brandt-1999a,Schuster-1994,Clem-2008,Morozov-1997}
Consequently, when the field $H_z$ is increased above the
penetration field $H_p\simeq (2H_{c1}/\pi)\sqrt{d/w}$
vortices entering through the edges are swept by the
Meissner currents toward the center where they accumulate
giving rise to a dome shaped induction profile of width
$2b$, see Fig. \ref{dome_dH}(c). The vortex-filled dome is
current free while the surrounding vortex-free region
carries the Meissner currents. The induction profile, with
a central dome surrounded by an area of zero $B$ is very
distinctive and is the hallmark of the GB model. The
vortices in the dome cannot leave the sample since they are
trapped by the barriers due to the Meissner currents in the
surrounding flux-free regions. Thus, when the field is
decreased the dome expands, while keeping the total flux in
it constant, giving rise to hysteretic magnetization. Only
when the dome reaches the edges the barriers vanish
allowing vortices to leave the sample.


\section{Experimental Details}

Several over-doped and optimally-doped BSCCO samples were
studied. We present here data of two over-doped samples,
with a critical temperature of $T_{c}=87$K, cleaved from
one thick crystal, sample A and sample B, with platelet
shapes of length $2L=1500\um$, width $2w=420 \um$ and
thicknesses $d$ of $8$ and $20\um$ respectively. Similar
results were obtained from optimally doped
crystals.\cite{Gutman-2009} The differential
magneto-optical (DMO) technique\cite{Soibel-2000} was used
to study the induction profiles and the magnetization
behavior of the samples. By minimizing the distance between
the magneto-optical indicator and the sample, and by
utilizing differential imaging we achieved high visibility
of the vortex chains and studied their effect on hysteretic
magnetization. The total exposure time of a differential
image was typically 2 to 3 minutes. We found that following
a field ramp the JV's often show slow relaxation for about
two minutes, during which their position within the sample
changed. Thus quality images of the chains could be
achieved by waiting 2 to 3 minutes between the end of a
field ramp and the beginning of image exposure. In order to
simulate the effect of JV's, we also studied the magnetic
behavior of crystal A after two fine grooves, 80 and 160 nm
deep and 350 nm wide, were etched on the surface using an
FEI Helios Focused Ion Beam system.


\section{results}

\begin{figure}[t]
 \centering \includegraphics[width=0.48\textwidth,
 bb=15 2 380 267, clip=,
 ]{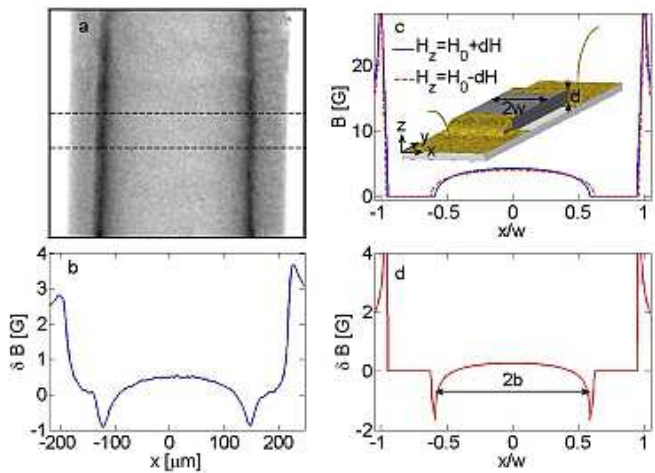}
\caption{(color online) (a) Field modulated DMO image of sample B
displaying the GB dome along the central part of the sample
($H_z=6.7$ Oe, $dH=0.4$ Oe, $H_x=0$, and $T=75$~K). The dark (white)
vertical stripes indicate the dome (sample) edges. (b) The
differential field profile $\delta B(x)$ across the width of the
sample averaged over the strip marked by the dashed lines in (a).
The two minima with negative $\delta B$ indicate the dome edges, and
the two maxima -- the sample edges. (c) The theoretical field profiles of a thin and infinitely long  sample $B(x,z=0)$, at two values of $H_z=H_0 \pm dH$ ($H_0=6.7$ Oe,
$dH=0.4$ Oe, $H_{c1}=33.1$ Oe, $d/w=0.1$). Inset: The experimental
schematics showing a BSCCO crystal with gold contacts glued onto a
substrate. (d) The difference $\delta B(x)$ of the two $B(x)$
profiles plotted in (c) corresponding to the experimentally measured
differential signal. The dome width $2b$ is marked.} \label{dome_dH}
\end{figure}

In order to visualize the vortex dome structure with
enhanced sensitivity, we carried out DMO with either field
modulation or current modulation. In the former case,
$B(x,y)$ images taken at $H_z=H_0-dH$ are subtracted from
images taken at $H_z=H_0+dH$ and few tens of such
differential images are averaged resulting in the
differential image $\delta B(x,y)$.\cite{Soibel-2000} When
the field is modulated by $2dH$ the vortex dome expands and
shrinks periodically, keeping the total flux in the dome
constant; see Fig. \ref{dome_dH}(c). As a result, when the
field is raised to $H_z=H_0+dH$ the induction in the center
of the dome is increased giving rise to a positive $\delta
B$ signal, while the induction near the dome edges is
decreased resulting in a negative $\delta B$
there,\cite{Morozov-1996,{Morozov-1997}}; see Fig.
\ref{dome_dH}(d). This effect is clearly visible in the DMO
data in Figs. \ref{dome_dH} (a) and (b), in which the edges
of the dome appear as dark regions of negative values.

\begin{figure}[t]
 \centering \includegraphics[width=0.5\textwidth%
 ]{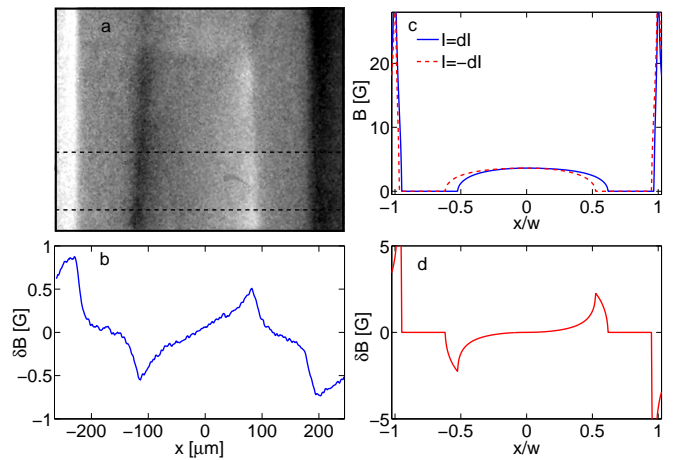}
\caption{(color online) (a) DMO image with current
modulation of sample B displaying the GB dome along the
sample center at $H_z=6.8$ Oe, $H_x=0$, $T=75$~K, and
$dI=10$ mA. The current modulation causes the right (left)
edge of the dome to appear bright (dark) and the right
(left) edge of the sample to appear dark (bright). (b) The
differential field profile across the width of the sample
averaged over the strip marked by the dashed lines in (a).
(c) Theoretical field profiles across  an infinitely long
and thin  sample in the presence of positive and negative
applied transport current showing the shift of the dome.
(d) The difference of the two profiles in (c) corresponding
to the experimentally measured differential signal.}
\label{dome_dI}
\end{figure}

In DMO with current modulation, images taken in the
presence of transport current $I=-dI$ are subtracted from
images taken at $I=dI$ and averaged.\cite{Banerjee-2004}
Since the Meissner currents flow in the opposite directions
on the two sides of the dome, the applied transport current
increases the total current flowing on one side of the dome
and reduces the current on the other side. As a result, the
dome is shifted away from the center as shown in Fig.
\ref{dome_dI}(c). Therefore, in the current modulated DMO
images one edge of the dome will appear as a bright strip
with a positive signal and the other as a dark strip with a
negative signal as shown theoretically in Fig.
\ref{dome_dI}(d) and experimentally in Figs. \ref{dome_dI}
(a) and (b). Note also that the self-induced magnetic field
of the transport current has an opposite sign at the two
edges of the sample. As a result in DMO image with current
modulation one edge of the sample is dark while the other
is bright, Fig. \ref{dome_dI}(a), in contrast to DMO with
field modulation in which both edges appear bright, Fig.
\ref{dome_dH}(a).

\begin{figure}[b]
 \centering \includegraphics[width=0.5\textwidth,
  bb=-225 60 900 500,clip=,
 ]{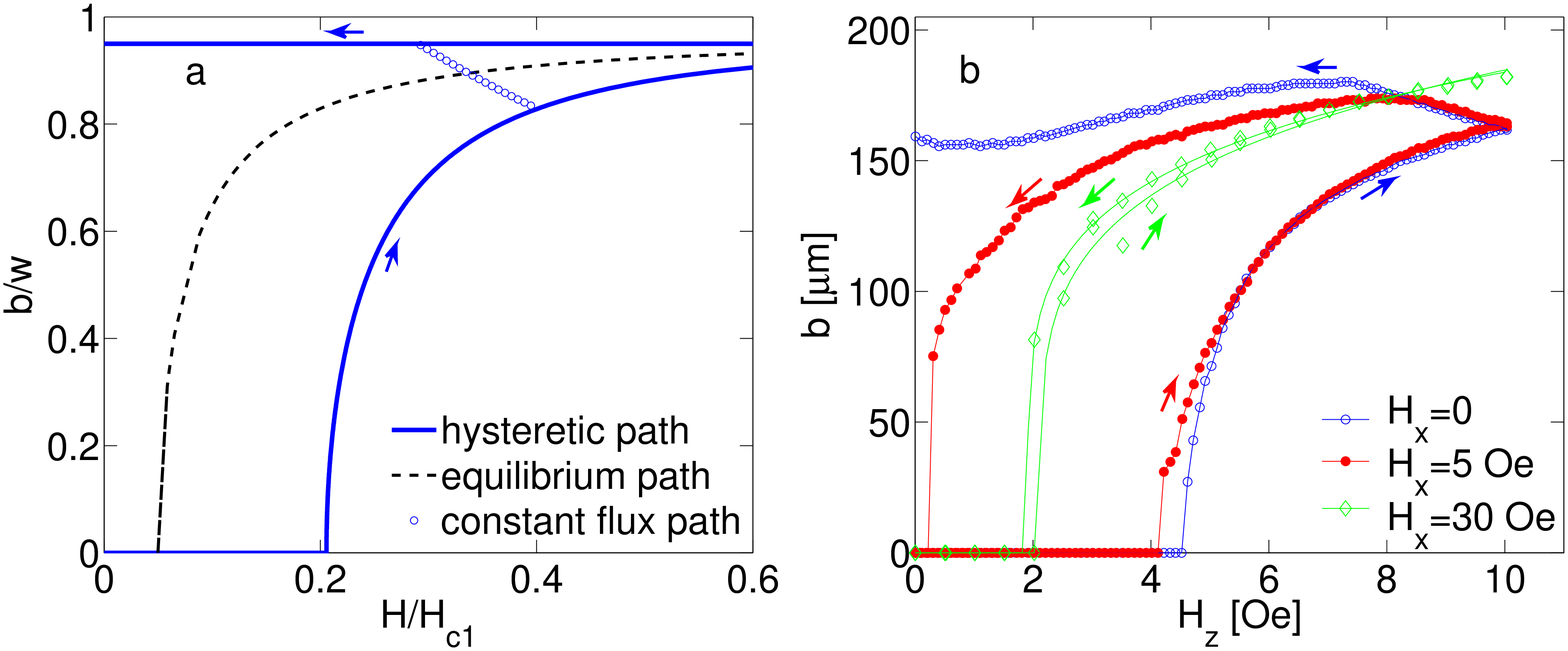}
\caption{(color online) (a) The calculated half-width of the vortex
dome $b$ as $H_z$ is increased from zero to $H_z \gtrsim H_{c1}$ and
decreased to zero again (solid line) within the GB model for an
infinitely long strip with $d/w=0.1$. When $H_z$ decreases from an
intermediate value, the dome widens following a constant flux path
(open circles) until it reaches the maximum value of  $b=w-d/2$ at
which the flux can exit the sample. The equilibrium dome width
(dashed line) is shown for comparison describing the conditions of
zero vortex energy in the dome. (b) The dome width measured in
sample B at $T=75$ K while ramping $H_z$ from zero to $10$ Oe and
back in the presence of zero and non zero $H_x$ showing that the
in-plane field equilibrates the GB hysteresis.} \label{bVsH}
\end{figure}

\begin{figure}[t]
 \centering \includegraphics[width=0.5\textwidth,
 bb=-235 50 900 490,clip=,
 ]{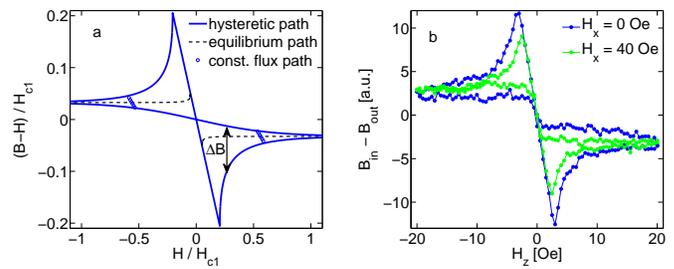}
\caption{(color online) (a) The calculated local magnetization loop
$B-H$ in the sample center within the GB model for the same
parameters as in Fig. \ref{bVsH}(a) (solid line) as compared to
equilibrium behavior (dashed). $\Delta B$ is the width of the
hysteresis. (b) The measured $B-H$ in the center of sample A (before
FIB etching) at $H_x=0$, $40$ Oe  and $T=79$ K.} \label{magloops}
\end{figure}

We now analyze the behavior of the dome width,  $2b$, along
the magnetization loop. Figure \ref{bVsH}(a) shows the
calculated behavior of $b$ using the GB formalism of Refs.
\onlinecite{Zeldov-1994,Morozov-1997,Zeldov-1994a}.
Initially the sample is in the Meissner state ($b=0$) which
is retained until $H_z=H_p\simeq (2H_{c1}/\pi) \sqrt{d/w}$,
above which flux starts entering the sample causing $b$ to
increase rapidly. As $H_z$ is further increased, $b$
asymptotically approaches its maximum value $b_{max}=
w-d/2$, corresponding to the dome width at which the
potential barrier vanishes. If the field is decreased from
an intermediate value, $b$ first expands to $b_{max}$ while
the total flux in the dome remains constant (open circles
in Fig. \ref{bVsH}(a)). When the field is further reduced,
$b$ remains at its maximal value allowing the vortices to
exit through the edges. The corresponding calculated local
magnetization loop is shown in Fig. \ref{magloops}(a).

It can be readily shown\cite{Zeldov-1994} that both the
ascending and descending magnetization branches are
energetically metastable. Compared to the equilibrium
conditions, at which the vortices in the dome have the same
energy as the vacuum $E=0$, on increasing (decreasing)
field the vortex energy is negative (positive), the dome is
narrower (wider), there is a deficiency (excess) of
vortices in the sample and the magnetization is lower
(higher) than at equilibrium. The dashed lines in Figs.
\ref{bVsH}(a) and \ref{magloops}(a) show the corresponding
equilibrium behavior of the dome width and of the local
magnetization.

Figure \ref{bVsH}(b) shows the experimentally measured dome
width along the magnetization loop in sample B. In the
absence of an in-plane field ($H_x=0$) the behavior is
qualitatively very similar to the theoretical GB curve
shown in Fig. \ref{bVsH}(a). On increasing $H_z$ the sample
is initially in the Meissner state ($b=0$) until
$H_z>H_p=4.5$ Oe where $b$ starts increasing rapidly. On
decreasing $H_z$ from its maximum value (10 Oe), $b$
continues to increase initially until it reaches a maximum
value, which is then roughly retained on further field
decrease.\footnote{The slight decrease in the value of $b$
on decreasing field can be readily attributed to small
remnant in-plane field considering the dramatic effect of
non-zero $H_x$, and to the finite $dH$ field modulation
used in DMO imaging.} This hysteretic behavior of the GB is
markedly altered by the presence of even a small in-plane
field. For $H_x=5$ Oe the penetration field $H_p$ is
slightly reduced to $4.1$ Oe and the ascending field branch
is similar to the $H_x=0$ case. However, on the descending
branch, $b$ is significantly below its maximum value and
decreases rapidly, reaching zero at finite $H_z$. The
shrinkage of the dome width on decreasing field is in sharp
contrast with GB model according to which vortices are
trapped in the dome as long as the surrounding vortex-free
region is present ($b<w-d/2$), and can leave the sample
only when dome reaches the sample edges ($b=w-d/2$). This
indicates that in the presence of in-plane field the
vortices can surprisingly exit the sample by permeating
through the vortex-free region that is impenetrable within
the GB model. At a higher in-plane field $H_x=30$ Oe the
dome is formed already at $H_z=2$ Oe, it shrinks back to
zero with decreasing field for $H_z<2$ Oe, and the dome
width $b$ behaves essentially reversibly following the
equilibrium curve (dashed) of Fig. \ref{bVsH}(a). This
shows that the vortices in the dome are effectively in
equilibrium with the vacuum and unexpectedly can now
readily move in and out of the dome through the vortex-free
region as $H_z$ is varied.

The DMO imaging allows a very sensitive measurement of the
dome width, but cannot provide a measure of the dc
magnetization. In order to analyze the sample magnetization
we have carried out a dc MO study in sample A. Due to the
lower sensitivity the resulting data in Fig.
\ref{magloops}(b) are nosier and less accurate, but
qualitatively consistent. Figure \ref{magloops}(a) shows
the calculated local magnetization loop $B-H$ in the center
of the sample (solid lines) along with the equilibrium
magnetization (dashed line). The large hysteresis is the
result of the GB that gives rise to wide flux free regions
on ascending field and essentially no flux-free regions on
descending field. The experimentally measured local
magnetization loop using MO imaging without modulation is
shown in Fig. \ref{magloops}(b). In the  absence of an
in-plane field ($H_{x}=0$) the behavior is consistent with
the GB model. Application of $H_{x}$, however,
significantly reduces the hysteresis on the ascending and
descending branches showing that $H_{x}$ facilitates both
vortex entry and exit.

\begin{figure}[b]
 \centering \includegraphics[width=0.5\textwidth,
 bb= 97 250 477 527, clip=,
 ]{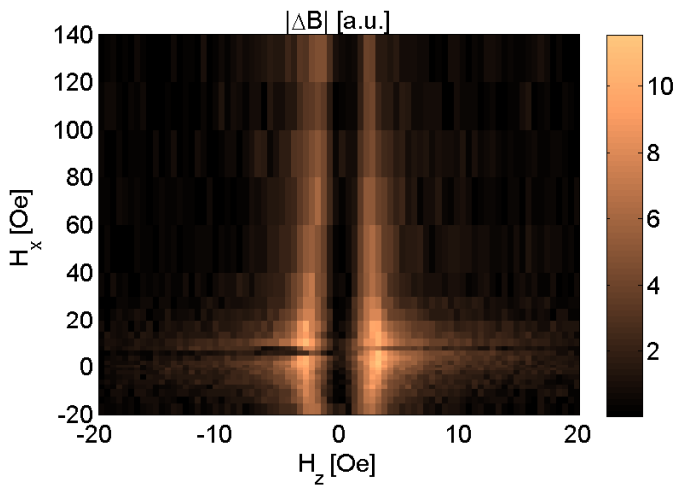}
\caption{(color online) The width of the magnetization loop,
$|\Delta B|$ (see Fig. \ref{magloops}), measured by sweeping $H_z$
at various values of $H_x$. The discontinuities about $H_x\approx 0$
are an artifact due to the MO indicator.} \label{dMpcolor}
\end{figure}

In order to quantify the reduction of the magnetic
hysteresis we plot in Fig. \ref{dMpcolor} the measured
hysteresis width, $|\Delta B|$ (see Fig.
\ref{magloops}(a)), as a function of $H_{z}$ and $H_{x}$.
The figure shows that as $|H_{x}|$ is increased the
hysteresis is suppressed significantly, however, the loop
does not close completely and some hysteresis is seen
around the penetration field, even at the highest in-plane
field we measured, 140 Oe.

\begin{figure}[t]
 \centering \includegraphics[width=0.5\textwidth,
 ]{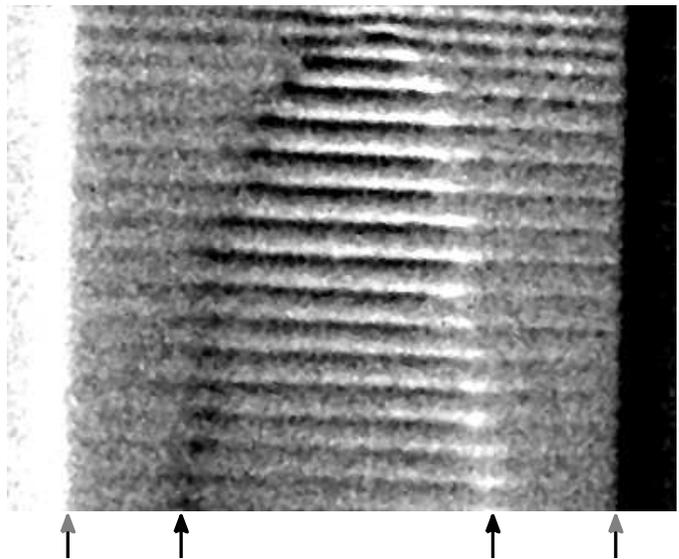}
\caption{A current modulated DMO image of the decorated JV's (slightly tilted horizontal lines) inside and outside the dome in sample B. The two side edges of the sample are marked by gray arrows and the dome edges by black arrows. The center of the sample coincides approximately with the image bottom while the top edge of the sample is slightly above the top of the image. As a result, the visible dome has  the shape of half an ellipse. The decorated JV's display a large contrast inside the dome while their whisker-like protrusions outside the dome display a lower contrast. The image was taken at $T=77$~K, $H_z=3.6$ Oe, $dI=6$ mA and the gray scale corresponds to about 0.5 G. The image width is 0.4 mm.} \label{chainIm}
\end{figure}

The evidence shown so far points to the fact that the JV's
introduced by the in-plane field significantly reduce the
effectiveness of the geometrical barriers for some reason.
In order to comprehend the microscopic origin of this
phenomenon we have optimized our DMO imaging technique to
resolve the PV chains along the JV's. The stacks of PV's
that decorate the stacks of JV's were previously observed
by Bitter decoration,\cite{Bolle-1991,Grigorieva-1995}
scanning Hall probes\cite{Grigorenko-2001,Grigorenko-2002}
and MO imaging.\cite{Vlasko-Vlasov-2002,Tokunaga-2003,
Tamegai-2004} We find that by adjusting the modulation
parameters the visibility of the chains can be greatly
enhanced using either field or current modulation. This is
demonstrated in Fig. \ref{chainIm}, where the vortex chains
are clearly visible as bright slightly tilted horizontal
lines that are aligned along the direction of the in-plane
field $H_x$. We verified that these lines are indeed vortex
chains decorating the stacks of JV's by rotating the
in-plane field and by varying $H_x$. We find that the lines
are always aligned along the direction of $H_x$ and their
separation follows $\sqrt{2\gamma \Phi_0/\sqrt{3}|H_x|}$
with $\gamma=480\pm 70$, where $\rm \Phi_0=20.7~G{\mu}m^2$
is the flux quantum and $\gamma$ is the anisotropy
parameter. In Fig. \ref{chainIm} the vortex dome is visible
as a half-oval envelope that has a bright contrast on the
right-hand-side and dark on the left-hand-side due to the
current modulated DMO. We expect PV's to be present only
inside the dome, residing on the JV stacks and between them
depending on field strength,\cite{Koshelev-1999} while in
the vortex-free regions outside the dome the PV's should be
absent due to the large Meissner currents and
correspondingly high vortex
potential.\cite{Zeldov-1994,Indenbom-1994,Benkraouda-1996,Brandt-1999,Brandt-1999a}
The striking observation in Fig. \ref{chainIm} is that the
vortex chains are visible both inside the dome as well as
outside the dome forming whisker-like protrusions that
extend all the way to the sample edges. This means that in
the presence of an in-plane field PV's can reside in the
regions outside the dome that are vortex free in the
absence of $H_x$. A similar situation was observed by
decoration experiments on BSCCO crystals in Ref.
\onlinecite{Tamegai-2004a}. In Fig. \ref{chainIm} the vortex
chains inside the dome have a stronger contrast as compared
to the chains outside the dome. We find that this is not
always the case and the whiskers may have stronger contrast
than the chains in the dome depending on the field and
modulation parameters.

\begin{figure}[t]
 \centering \includegraphics[width=0.48\textwidth,
 bb=-130 60 560 710, clip=,
 ]{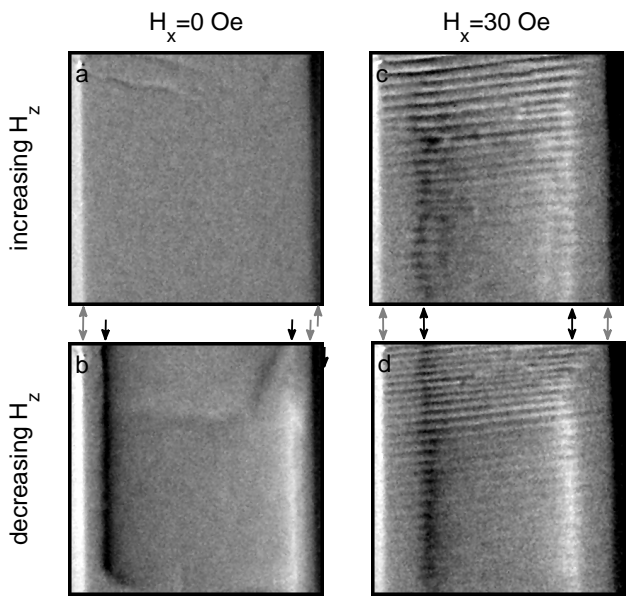}
\caption{Current modulated DMO images at $H_z=3$ Oe ($dI=10$ mA, $T=75$ K) of sample B taken along the magnetization loops shown in Fig \ref{bVsH}(b) for $H_x=0$ (left column) and $H_x=30$ Oe (right column) on ascending (top row) and descending (bottom row) $H_z$ branches. The left edge of the dome is dark and the right is bright (marked by black arrows), while the left edge of the sample is bright and the right is dark (marked by gray arrows). The slightly tilted horizontal stripes in (c) and (d) are PV chains. The gray scale is about 2.5 G in (a) and (b) and 1.5 G in (c) and (d). The images are 0.4 mm wide.} \label{dI4ims}
\end{figure}

\begin{figure}[t]
 \centering \includegraphics[width=0.48\textwidth,
 bb=-130 60 560 710, clip=,
 ]{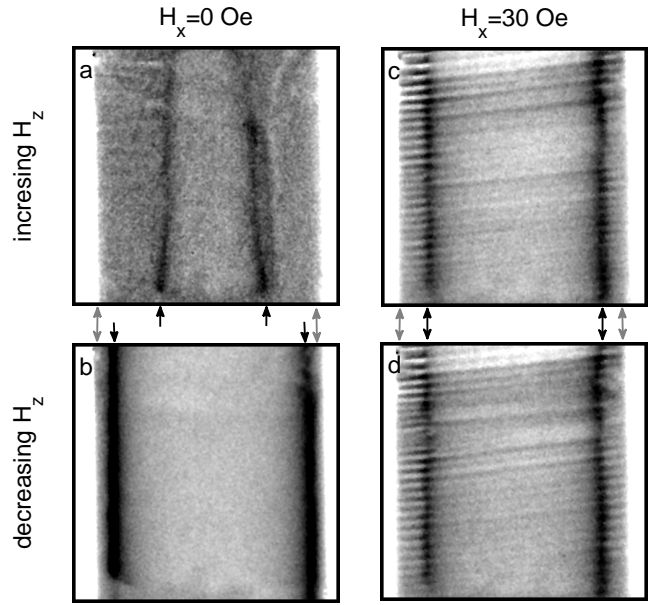}
\caption{Field modulated DMO images at $H_z=5$ Oe ($dH=0.4$
Oe, $T=75$ K) of sample B taken along magnetization loops
in the presence of $H_x=0$ (left column) and $H_x=30$ Oe
(right column) on ascending (top row) and descending
(bottom row) $H_z$ branches. The dome edges are dark
(marked by black arrows) and sample edges are bright
(marked by gray arrows). The slightly tilted horizontal
stripes in (c) and (d) are PV chains. The gray scale is
about 1.25 G, 2.5 G, 1.8 G and 1.9 G in (a) to (d)
respectively. The images are 0.5 mm wide. A movie showing
the complete magnetization loops is available in Ref.
\onlinecite{AuxWebPage}.} \label{dH4ims}
\end{figure}

We  now present the hysteretic behavior of the dome with
and without $H_x$. Figure \ref{dI4ims}(a) shows a DMO image
using current modulation at $H_z=3$ Oe in the absence of an
in-plane field. At this field the sample is in the Meissner
state on the ascending branch and the vortex dome is
absent, since $H_z<H_{p}$. On the descending branch,
however, the sample is in the mixed state and the dome is
maximal as shown in Fig. \ref{dI4ims}(b). This highly
hysteretic behavior of the dome is the result of flux
trapping due to GB. In contrast, in the presence of
$H_x=30$~Oe the dome has the same width on the ascending
and  the descending fields, showing that the vortices are
in equilibrium state. Figure \ref{dH4ims} shows similar DMO
images using field modulation at $H_z=5$~Oe. In absence of
an in-plane field Fig. \ref{dH4ims}(a) shows a narrow dome
just above the penetration on ascending field, while Fig.
\ref{dH4ims}(b) shows a very wide dome on the descending
$H_z$ branch. When $H_x=30$~Oe is applied, the domes on
ascending and descending branches become essentially
identical. A movie showing the complete magnetization loop
from which the \fig{dH4ims} is extracted is available in
Ref. \onlinecite{AuxWebPage}.

The energy of the PV's that reside on JV's is lower due to
their attractive interaction. Since the JV's transverse the
entire sample, one could argue that the energy of the PV's
could be lowered sufficiently by the JV stacks  to extend
into the GB in the vortex free region. The GB energy scale
is $E_{gb}=\eps_0d$, where $\eps_0=H_{c1}\Phi_0/4\pi$, with
the barrier being highest near the sample edges and
gradually decreasing toward the dome.\cite{Zeldov-1994} The
JV-PV interaction strength per unit length of the PV stack,
is given by crossing energy\cite{Koshelev-1999}
\begin{equation}
\label{Ex}
\eps_\times\approx\sqrt{\frac{\sqrt{3}\gamma B_x}{2\Phi_0}}\frac{2.1\Phi_0^2}{4\pi^2\gamma^2s
\ln(3.5\gamma s/\lambda)},
\end{equation}
where $\gamma=480$ as described above, $s=15${\AA} {is} the
layer spacing, and $\lambda$ is the London penetration
depth such that $\lambda=\lambda(0)/\sqrt{1-(T/T_c)^2}$
with $\lambda(0)=2000${\AA}. In the case of our
experimental conditions the resulting total vortex energy
reduction is $\eps_\times d \sim 10^{-2}E_{gb}$. Therefore,
the degree of the spacial extension of the PV whiskers into
the vortex-free region should be extremely small,
$\mathcal{O}(10^{-2})$ of the width of the flux-free
region, and hence should be hardy observable. Nevertheless,
our images clearly show that PV chains extend through the
entire vortex-free region, in sharp contrast to the above
estimates.

\begin{figure}[t]
 \centering \includegraphics[width=0.4\textwidth,
 bb=130 164 508 650, clip=,
 ]{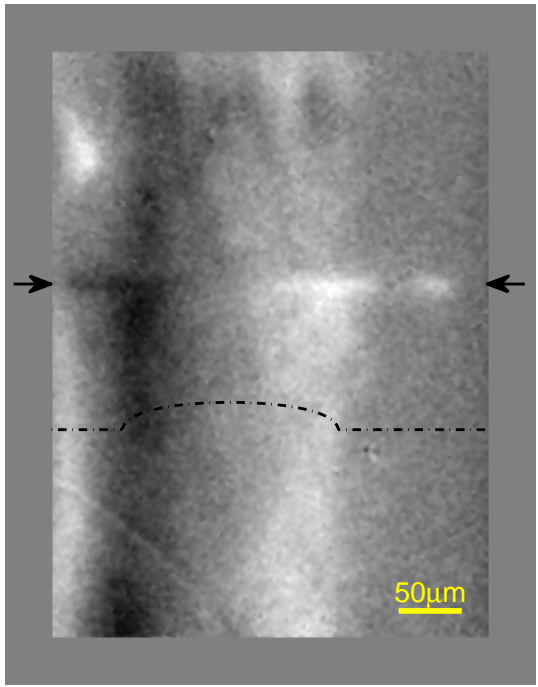}
\caption{(color online) A zoomed-in view of the vortex dome in the central part of sample A after a groove 160 nm deep and 350 nm wide was etched across the width of the sample using FIB (marked by arrows). The image was obtained using current modulated DMO on ascending $H_{z}=4$ Oe at $T=82$~K and $dI=10$ mA. The dome left (right) edge is dark (bright) and its schematic induction profile is shown by the dot-dashed line. Vortex whiskers protruding outside the dome are visible along the groove.} \label{groove}
\end{figure}

It  was recently shown that small local inhomogeneities in
$H_{c1}$ could lead to very large variations in the
equilibrium local induction, a phenomenon attributed to
demagnetization effects.\cite{Avraham-2008} In order to see
whether demagnetization can explain the phenomenon at hand
we studied the GB mechanism in sample A after two shallow
grooves were etched on its surface across the width of the
sample as described above. We found that the grooves have a
surprisingly large effect: a 160 nm deep groove, just $2\%$
of the sample thickness, reduces the GB hysteresis
comparable to the suppression by the in-plane field of
about 10 Oe, and vortex whiskers are formed along the
groove. The current modulated DMO image in Fig.
\ref{groove} shows that on ascending field the whiskers
along the groove (marked by arrows)  have a contrast that
is comparable to the dome in the pristine region but their
extent is about $60\%$ wider than the pristine dome. On the
descending $H_{z}$ branch the vortex dome in the pristine
region closes gradually while the whiskers along the groove
extend up to the sample edges similar  to the behavior of
the whiskers along the JV's.

\section{discussion}

The reduction of the GB by JV stacks has four main
manifestations in our data: (\textit{i}) suppression of the
magnetic hysteresis, (\textit{ii}) reduction of the
penetration field with increasing $|H_x|$, (\textit{iii})
the presence of PV chains outside the bulk vortex dome, and
(\textit{iv}) on the descending $H_z$ branch of the
magnetization loop---the reduction of the bulk dome width
to zero at some finite $H_z=H_0$, with $H_0$ increasing
with increasing $|H_x|$. A shallow and narrow groove in the
absence of $H_x$, had the same qualitative effects on the
GB. Indeed, both the groove and a stack of JVs reduce the
energy of the vortices residing on them: A groove reduces
the average vortex energy by $\eps_0\delta/d$, where
$\delta$ is the groove depth, while a stack of JVs reduces
the vortex energy by $\eps_\times$. Thus in both cases a
chain-like structure with enhanced density of vortices is
expected to be formed along the groove or the JVs. In the
following discussion, we therefore use the term
\emph{groove} in order to describe the related reduction of
the geometrical barrier in both experimental situations.
Note that in contrast to the fixed groove fabricated by
FIB, in the case of JVs both the density of the ``grooves''
and their effective ``depth'' $d\eps_\times/\eps_0$
increase with $H_x$ (see Eq. \eqref{Ex}). As a result, in
the case of JV ``grooves'' one may expect a stronger
suppression of the GB with increasing $H_x$ as observed
experimentally.

In the discussion that follows we show that the suppression
of the vortex energy along the groove is insufficient to
account for the observed phenomena on its own. We then
demonstrate that the boundary conditions require the
presence of substantial shielding currents that flow along
the groove. These currents cause an additional significant
increase of the magnetic induction inside the groove
enhancing the suppression of the GB in accordance with the
experimental results.

\begin{figure}[b]
 \centering \includegraphics[
 width=0.5\textwidth,
 bb=-71 115 680 450, clip=,
 ]{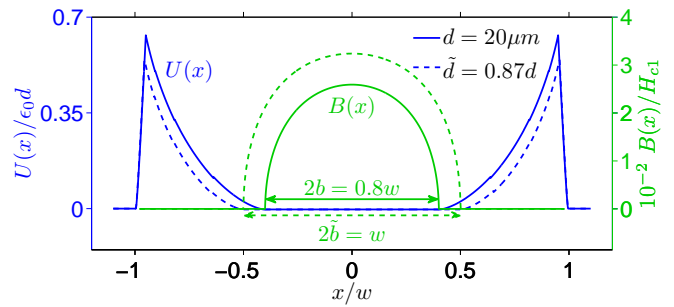}
\caption{(color online) The equilibrium potential curves
(blue), $U(x=0)=0$, and induction profiles (green) for two
regions of a superconductor having different thicknesses,
$d$ (continuous line) and $\tild$ (dashed line), assuming
both regions can be described by the GB mechanism. The
difference in the edge barrier height in the two regions
which is proportional to $\delta/d=(d-\tild)/d$, is of the
same order of magnitude as the {normalized} dome width
difference $(\tilde{b}-b)/b$.
}
 \label{Ecomparison}
\end{figure}

We start by evaluating the expected reduction of the GB due
to a groove across the width of the sample. Let us consider
a sample of width $2w$ and thickness $d$ ($-d/2\leq z \leq
d/2$) having a groove of depth $\delta$ extending from
$x=-w$ to $x=w$. Since the vortex energy in the groove is
lower, we expect that the vortex dome along the groove will
be wider than in the bulk forming whisker-like protrusions
of vortices into the flux free region. If the groove width,
denoted by $2l$, is large enough, i.e. $l\gtrsim w$ then it
can be described by the GB formulation following for
example Ref. \onlinecite{Zeldov-1994a}. Figure
\ref{Ecomparison} shows for example the resulting vortex
potential $U(x)$ and field $B(x)$ profiles in the bulk and
along the groove of depth $\delta = 0.13 d$ at $H_z=0.065
H_{c1}$ under equilibrium conditions, $U(x=0)=0$. Here the
dome width in the bulk is $b=0.4w$ while the dome in the
groove has $\tilde{b}=0.5w$ resulting in whiskers that
extend out of the bulk dome by $25\%$. Experimentally, in
contrast, both the FIB etched groove and the effective
grooves along the JV stacks have a much stronger effect:
the FIB groove in Fig. \ref{groove} shows similar extent of
the whiskers for groove depth of just $\delta = 0.02 d$,
whereas the JV stacks result in whiskers that commonly
extend all the way to the sample edges, with effective
groove depth of the order of just $\delta = 0.01 d$.

\begin{figure}[t]
 \centering \includegraphics
[width=0.45\textwidth,bb=53 252 528 510,clip=,]
 {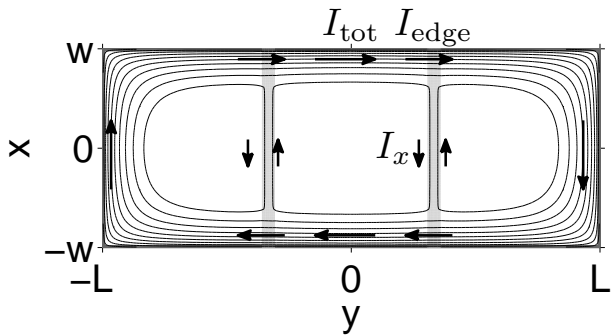}
\caption{A numerical solution for the sheet current stream
lines of a two-dimensional rectangular film of half-length
$L$ and half-width $W$ ($W/L=0.4$) with two grooves (shaded
regions), each $2l=0.06L$ wide, at $H_z/H_{c1}=0.44$. The
arrows indicate schematically the current flow directions.
It is assumed that the penetration of the vortices into the
groove begins when the sheet Meissner current at a distance
$d/2$ from the platelet edge reaches $cH_{c1}/2\pi$. Here
we took $d=0.05L$. }
\label{Jcontour}
\end{figure}

The  above calculation, however, does not take into account
the current $I_x$ that flows along the groove-bulk
interface, see \fig{Jcontour}, which may significantly
enhance the described effect. This current plays a similar
role to that of the edge current in the GB model and may
significantly enhance the magnetic induction in the groove.
Vortices can penetrate from the groove into the bulk only
if the difference in their line energy $\epsilon_0\delta$
is canceled by the Lorentz potential due to that part of
$I_x$ which flows in a region of width $\sim \delta$ near
the groove,
 \begin{equation}
\label{j=delta}
  \frac{\Phi_0}{c} \int_l^{l+\delta} J_x(y)dy=\epsilon_0\delta,
 \end{equation}
where $J_x(y)$ is the appropriate sheet current. $J_x(y)$
always keeps the induction in the groove  higher than in
the bulk, thus enhancing the effect of the groove and
increasing the extent of the vortex whiskers. In the
Meissner state $J_x(y)=0$ and no current flows along the
groove edges. At low fields, when the vortices are present
only in the groove, $J_x(y)$ extends over the entire bulk.
With increasing field, dome is formed in the bulk,
decreasing $J_x(y)$. When the field is high enough so $J_x$
flows only in the region of width $\sim \delta$ at the
interface, the \emph{total} interface current $I_x$ reaches
its minimum value $I_x^0=c\epsilon_0\delta/\Phi_0 =c
H_{c1}\delta/4\pi$ given by Eq. \eqref{j=delta}. The
difference in the inductions at the groove center and bulk
$\Delta B_{gr}$ which is proportional to $I_x$, depends on
the geometry of the groove. For example, if the whole
current $I_x$ flows near the groove, one has  $\Delta
B_{gr}= (4\pi/c) I_x/d$ for the narrow groove, $l\ll d$,
and $\Delta B_{gr}= (4/c) I_x/l$ in the opposite limiting
case, $l\gg d$. Note that the enhancement of the field in
the groove is essentially a demagnetization effect due to
partial expulsion of the flux from the bulk. It is clear
that for very wide grooves the effect is negligible and the
result of Fig. \ref{Ecomparison} holds.

\begin{figure}[b]
 \centering \includegraphics
[width=0.5\textwidth]
 {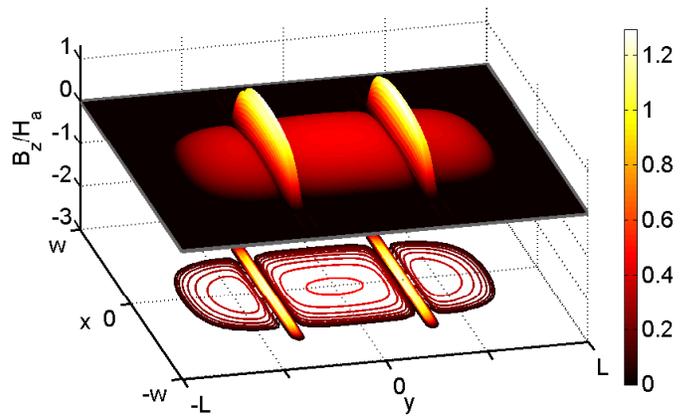}
\caption{(color online) A numerical solution for magnetic induction $B_z$ corresponding to the conditions of Fig.
\ref{Jcontour}. The field in the two grooves is higher than the applied field similar to the
situation described in Ref \onlinecite{Avraham-2008}.} \label{B2slits}
\end{figure}

The  sheet current ${\bf J}(x,y)$ and magnetic induction
$B_z(x,y)$ can be calculated numerically for a finite size,
two-dimensional film using the method described in
Ref.~\onlinecite{Brandt-2005a}. This method can be applied
to the problem at hand in the limit of $d\ll l \ll w$. Here
we demonstrate by this calculation that the dome in the
groove can be noticeably wider than the dome in the bulk.
To carry out the calculation, we rewrite condition
\eqref{j=delta} using the known solution for the sheet
current and magnetic induction in the case when there is a
magnetic flux in a narrow slit of the width $2l$ between
two wide infinitely long strips in the Meissner
state.\cite{Brojeny-2002} This solution can be used only if
$J_x$ flows in the region essentially wider than $d$, i.e.,
if $(4\pi/c)I_x \gg H_{c1}\delta$. Then,
Eq.~(\ref{j=delta}) can be rewritten as
\begin{equation}
\label{Bz=AHc1}
B_z(x=0,y=0)=AH_{c1}, 
\end{equation}
where $A$ is a factor  which depends on the distribution of
the current density in the layer of the thickness $\delta$
near the groove. Strictly speaking, this distribution is
essentially three-dimensional and cannot be found from the
solution of Ref. \onlinecite{Brojeny-2002}. We estimate that the factor $A$ lies in the interval from $\sqrt{\delta/l}$ to $\sqrt{d/l}$. Since the calculation is done for a strictly
2D film, $\delta$ enters the calculation only through $A$
which we take as $A=0.5$ in the figures shown next. Figures
\ref{Jcontour} and \ref{B2slits} show the current stream
lines and magnetic induction of a 2D superconductor having
two grooves of width $2l=0.06L$ at $H_z/H_{c1}=0.44$. The
interface current $I_x$ as well as the flux focusing into
the slits is clearly seen. Figure \ref{B2slits} shows that
the dome in the grooves is wider than the bulk dome forming
whiskers protruding into the flux free region. The
induction in the grooves is much higher than in the bulk
and interestingly, it is even higher than the applied
field. The paramagnetic response of the grooves, $B_z>H_z$,
is similar to the situation described in Ref.
\onlinecite{Avraham-2008}, where such a phenomenon was
described in terms of demagnetization effect.

Our numerical  calculations also provide a qualitative
understanding of the behavior on the descending field as
follows. When the applied field is decreased, the bulk dome
and the whiskers along the grooves expand until the
whiskers reach the sample edges. Since the bulk and groove
vortices remain in equilibrium, on further decrease of
$H_z$ the vortices are drained out from the bulk through
the groove. This process results in reduction of the width
of the hysteresis loop and in magnetization value that is
closer to the equilibrium magnetization on the descending
branch. The inductions at both the groove and bulk centers
decrease with decreasing $H_z$, but their difference,
$\Delta B_{gr}\propto I_x^0= cH_{c1}\delta/4\pi$ remains
constant. Thus as $H_z$ is reduced to zero the bulk dome
contracts until all its vortices are drained out at a field
$H_0=H_{b=0}$ such that the induction at the groove center
equals $\sim \Delta B_{gr}$. In order to estimate $H_0$, we
note that at this field $B_z=0$ in the bulk and the sample
is in the Meissner state everywhere except in the grooves.
In the Meissner state the total screening current
circulating in a strip is $I_{\rm
tot}=cH_0w/2\pi$.\cite{Larkin-1971a, Huebener-1972} Near
the grooves this current splits into two paths: $I_x^0$
flows parallel to the groove and $I_{\rm edge}$ flows
across the groove edge (see \fig{Jcontour}). Thus, we
obtain $H_0\approx H_{c1}\delta/2w + 2\pi I_{\rm edge}/cw$.
We can thus evaluate $H_0$ by noting that the edge current
lies in the interval $0\le I_{\rm edge} \lesssim c
H_{c1}d/4\pi$.\cite{Zeldov-1994} This estimate for the
range of $H_0$ is in  reasonable agreement with the
experimental values of $H_0$ seen on the descending
branches of $b(H_z)$ in Fig.~\ref{bVsH}b.

The above description provides a  qualitative explanation
of the creation of vortex whiskers, the suppression of
hysteresis, and the shrinkage of the bulk vortex dome on
descending field. Quantitatively, however, the groove
depths that are required in order to describe the
experimental data are still about an order of magnitude
larger than the ones obtained from Eq. \eqref{Ex}. Our
numerical calculations are applicable only for grooves that
are not too narrow $d\ll l \ll w$, while the JV stacks
create effective grooves with $l \ll d$. In this case we
expect further enhancement of the demagnetization effects
that could perhaps resolve the discrepancy. Interestingly,
the JV-PV crossing energy $\eps_\times$ was evaluated
experimentally to be about seven times larger than the
theoretical value by comparing $\eps_\times$ with the
vortex line energy in wide grooves.\cite{Tamegai-2004} The
origin of this discrepancy could be related to the same
observation that the effective grooves due to JV's are much
narrower, thus having a large field enhancement due to
demagnetization effects in the thin samples in
perpendicular field.

The  presented model, however, does not explain why the
penetration field decreases significantly with increasing
$H_x$. In the full Meissner state $I_x=0$ and therefore
$H_p$ cannot be affected by it. Since in the GB model
$H_p\simeq (2H_{c1}/\pi)\sqrt{d/w}$, the bare reduction of
the thickness along the JV effective grooves should lower
$H_p$ in our case by less than $1\%$, in sharp contrast to
more than $50\%$ reduction in $H_p$ in Fig. \ref{bVsH}(b).
This essential reduction of $H_p$ may be due to a more
intriguing mechanism in the vicinity of the edges in which
an inhomogeneous penetration of the vortices into the edge
region leads to a concentration of the magnetic field in
the groove edges. As a result, an enhanced penetration of
the vortices along the groove develops that noticeably
decreases the observed penetration field.


\section{Summary}
We used the DMO technique to image the field  distribution
and the local magnetization of thin BSCCO crystals at
elevated temperatures at which the magnetic hysteresis is
governed by the geometrical barrier mechanism. In this
regime a vortex dome is present in the central part of the
sample surrounded by a vortex-free potential barrier
region. We found that stacks of Josephson vortices caused
by an in-plane magnetic field significantly reduce the GB
and allow formation of pancake vortex chains or whiskers in
regions that should otherwise be vortex-free. These
whiskers extend from the dome up to the sample edges, thus
forming easy channels for vortex flow through the potential
barrier. As a result the magnetic hysteresis is reduced and
the magnetization loops become reversible at elevated
in-plane fields. We also found that a FIB etched narrow and
shallow groove has a similar effect on the GB.

Our analysis shows that the bare reduction of  the vortex
energy due to the JV-PV crossing energy along the JV stacks
or due to reduced thickness along the etched grooves is one
to two orders of magnitude lower than the height of the GB,
and therefore cannot account for the observed phenomena. We
present a model in which the effect of the grooves is
significantly enhanced due to the demagnetization factor in
the platelet geometry in perpendicular field. The flux that
is partially expelled from the pristine regions is focused
into the grooves, thus enhancing the local induction in the
grooves and suppressing the barrier. The presented
numerical calculations demonstrate qualitatively this
demagnetization enhancement, and provide an important
insight into the long-standing puzzle of suppression of the
macroscopic GB by in-plane field. Our numerical method can
only treat grooves that are wide on the scale of sample
thickness, which are much wider than in the experimental
situation. A proper calculation of narrower grooves is
therefore required, and is expected to provide a better
quantitative agreement with the data. One important
experimental observation, the suppression of the
penetration field by the JV stacks, however, remains
unexplained within the model. This feature is apparently a
result of microscopic inhomogeneities in the sample edge
regions introduced by the JV's that facilitate vortex
penetration into the sample. Such inhomogeneities cannot be
treated within our current numerical studies and will be
the subject of future investigations.

\begin{acknowledgments}
We wish to thank the Electron Microscopy Unit in  the
Weizmann Institute of Science for providing us access to
the FEI Helios Focused Ion Beam system. This research was
supported by the German-Israeli Foundation for Scientific
Research and Development (GIF) and by the US-Israel
Binational Science Foundation (BSF). EZ acknowledges the
support of EU-FP7-ERC-AdG.

\end{acknowledgments}
\bibliography{all2}
\end{document}